\documentclass[twocolumn,tighten]{aastex63}

\pdfoutput=1
\usepackage[utf8]{inputenc}
\usepackage[T1]{fontenc}
\usepackage{amssymb}
\usepackage{apjfonts}
\usepackage[figure,figure*]{hypcap}

\begin{document}

\title{Discovery of the Remarkably Red L/T Transition Object VHS J183135.58-551355.9}

\correspondingauthor{T. Bickle}
\email{tombickleastro@outlook.com}

\author[0000-0003-2235-761X]{Thomas P. Bickle}
\affil {School of Physical Sciences, The Open University, Milton Keynes, MK7 6AA, UK}

\author[0000-0002-6294-5937]{Adam C. Schneider}
\affil{United States Naval Observatory, Flagstaff Station, 10391 West Naval Observatory Rd., Flagstaff, AZ 86005, USA}

\author[0000-0002-2592-9612]{Jonathan Gagn\'e}
\affil{Plan\'etarium Rio Tinto Alcan, Espace pour la Vie, 4801 av. Pierre-de Coubertin, Montr\'eal, Qu\'ebec, Canada}
\affil{Institute for Research on Exoplanets, Universit\'e de Montr\'eal, D\'epartement de Physique, C.P.~6128 Succ. Centre-ville, Montr\'eal, QC H3C~3J7, Canada}

\author[0000-0001-6251-0573]{Jacqueline K. Faherty}
\affil{Department of Astrophysics, American Museum of Natural History, Central Park West at 79th Street, NY 10024, USA}

\author[0000-0003-4083-9962]{Austin Rothermich}
\affil{Department of Astrophysics, American Museum of Natural History, Central Park West at 79th Street, NY 10024, USA}

\author[0000-0003-0489-1528]{Johanna M. Vos}
\affil{School of Physics, Trinity College Dublin, The University of Dublin, Dublin 2, Ireland}
\affil{Department of Astrophysics, American Museum of Natural History, Central Park West at 79th Street, NY 10024, USA}

\author[0000-0002-2011-4924]{Genaro Su\'arez}
\affil{Department of Astrophysics, American Museum of Natural History, Central Park West at 79th Street, NY 10024, USA}

\author[0000-0003-4269-260X]{J. Davy Kirkpatrick}
\affil{IPAC, Mail Code 100-22, Caltech, 1200 E. California Blvd., Pasadena, CA 91125, USA}

\author[0000-0002-1125-7384]{Aaron M. Meisner}
\affil{NSF’s National Optical-Infrared Astronomy Research Laboratory, 950 N. Cherry Ave., Tucson, AZ 85719, USA}

\author[0000-0002-2387-5489]{Marc J. Kuchner}
\affil{Exoplanets and Stellar Astrophysics Laboratory, NASA Goddard Space Flight Center, 8800 Greenbelt Road, Greenbelt, MD 20771, USA}

\author[0000-0002-6523-9536]{Adam J. Burgasser}
\affil{Department of Astronomy \& Astrophysics, University of California San Diego, La Jolla, CA 92093, USA}

\author[0000-0001-7519-1700]{Federico Marocco}
\affiliation{IPAC, Mail Code 100-22, Caltech, 1200 E. California Blvd., Pasadena, CA 91125, USA}

\author[0000-0003-2478-0120]{Sarah L. Casewell}
\affil{School of Physics and Astronomy, University of Leicester, Leicester, LE1 7RH, UK}

\author[0000-0001-7896-5791]{Dan Caselden}
\affil{Department of Astrophysics, American Museum of Natural History, Central Park West at 79th Street, NY 10024, USA}

\author[0000-0001-8170-7072]{Daniella C. Bardalez Gagliuffi}
\affil{Department of Physics \& Astronomy, Amherst College, 25 East Drive, Amherst, MA 01003, USA}

\author{The Backyard Worlds: Planet 9 Collaboration}

\begin{abstract}

We present the discovery of VHS J183135.58$-$551355.9 (hereafter VHS J1831$-$5513), an L/T transition dwarf identified as a result of its unusually red near-infrared colors ($J-K_{\rm S}=3.633\pm0.277$ mag; $J-W2=6.249\pm0.245$ mag) from the VISTA Hemisphere Survey and CatWISE2020 surveys. We obtain low resolution near-infrared spectroscopy of VHS J1831$-$5513 using Magellan/FIRE to confirm its extremely red nature and assess features sensitive to surface gravity (i.e., youth). Its near-infrared spectrum shows multiple CH$_{\rm 4}$ absorption features, indicating an exceptionally low effective temperature for its spectral type. Based on proper motion measurements from CatWISE2020 and a photometric distance derived from its $K_{\rm S}$-band magnitude, we find that VHS J1831$-$5513 is a likely ($\sim$85$\%$ probability) kinematic member of the $\beta$ Pictoris moving group. Future radial velocity and trigonometric parallax measurements will clarify such membership. Follow-up mid-infrared or higher resolution near-infrared spectroscopy of this object will allow for further investigation as to the cause(s) of its redness, such as youth, clouds, and viewing geometry.

\end{abstract}

\keywords{Brown Dwarfs; L Dwarfs; T Dwarfs; Stellar Associations; Moving Clusters}

\section{Introduction}
At young ages, brown dwarfs have not yet contracted to their final equilibrium radii \citep{Burrows1997} and have low masses for their effective temperatures. They therefore have lower surface gravities and lower pressure atmospheres than their field-age counterparts. This low pressure reduces collision-induced absorption by H$_2$ \citep{Linsky1969,Borysow1997} and it is hypothesized that it causes an excess of clouds and dust in their photospheres \citep{Cushing2008,Faherty2016}, shifting emergent flux to longer wavelengths (e.g., \citealt{Faherty12}). Consequently, the near-infrared colors of young brown dwarfs are notably redder than those of field age objects of the same spectral type \citep{Faherty2016,Liu2016}.

Young, red L and T dwarfs can act as valuable analogs for young, giant exoplanets, possessing similar effective temperatures, masses, and atmospheric properties \citep{Liu2013,Faherty2016}. The planets HR8799bcde \citep{Marois2008} and planetary-mass companions VHS 1256$-$1257 b \citep{Gauza2015}, 2M1207b \citep{Chauvin2004} and BD+60 1417b \citep{Faherty2021} are apt examples of this, all exhibiting near-infrared spectra consistent with red, low-gravity, late-L dwarfs. Crucially, young free-floating substellar objects can be studied without the interference of light from a host star, providing convenient and effective laboratories for tests of planetary theory \citep{Faherty2016}. This convenience does, however, come at a cost; whereas companions can have properties (e.g., age and metallicity) inferred from their host star under the assumption of common formation, the properties of free-floating brown dwarfs are notoriously difficult to determine. An age-mass-temperature degeneracy exists for brown dwarfs \citep{Burrows1997}, making it challenging to pinpoint these three key properties for isolated brown dwarfs. However, if a brown dwarf can be linked through kinematics and age diagnostics to a known moving group with a well constrained age, this degeneracy inherent in brown dwarfs can be broken, and properties consistent throughout that group's membership, such as age (e.g., \citealt{Bell2015}) and metallicity (e.g., \citealt{Viana2009}), may be attributed to the object. When the age of an object is inferred, evolutionary models can be used to estimate physical properties such as mass and effective temperature (e.g., \citealt{Filippazzo2015,Suarez2021}). Such objects are vital benchmarks for studying brown dwarf evolution, and at young ages and low masses, can provide important empirical constraints to the initial mass function \citep{Gagne2017,Kirkpatrick2021,Kirkpatrick2024}.

While targeted searches using the Two Micron All-Sky Survey (2MASS; \citealt{Skrutskie2006}) have discovered unusually red, nearby L dwarfs \citep{Kellogg2015,Schneider2017}, the detection limit of 2MASS ($J\approx17$ mag) means that many low mass, red L/T dwarfs are missed. This restriction is the primary reason that the number of spectroscopically confirmed free-floating L/T dwarfs with extremely red colors ($J-K>2.2$ mag) remains low ($\sim$15). The advent of a new generation of surveys, such as the VISTA Hemisphere Survey (VHS; \citealt{mcmahon2013}), the UKIDSS Large Area Survey (UKIDSS LAS; \citealt{Lawrence2007}) and UKIRT Hemisphere Survey (UHS; \citealt{Dye2018}), which probe significantly deeper than 2MASS ($5\sigma$ \textit{J}-band depths of 20.2, 19.9 and 19.6 mag respectively), has brought the potential for the discovery of objects with fainter $J$-band magnitudes than previously possible (e.g., \citealt{Schneider2023}). Consequently, these surveys are ideal hunting grounds for young, red L and T dwarfs, the $J$-bands of which are inherently suppressed. Reaching fainter $J$-band magnitudes also means that redder objects than previously known have become detectable. Indeed, until \cite{Schneider2023}'s discovery of the reddest known free-floating L/T dwarf CWISE J050626.96+073842.4 (CWISE J0506+0738; $(J-K)_{\rm MKO}$=2.974$\pm0.039$ mag) using UHS, it had been approximately a decade since the discovery of the previously reddest known objects PSO J318.5338$-$22.8603 ($(J-K)_{\rm MKO}$=2.74$\pm0.04$ mag; \citealt{Liu2013}) and ULAS J222711$-$004547 ($(J-K)_{\rm MKO}$=2.79$\pm0.06$ mag; \citealt{Marocco2014}). The discovery of such extraordinarily red objects is important as it allows us to explore parameters that influence the appearance of brown dwarfs such as youth, metallicity, clouds, and viewing inclination (e.g., \citealt{Looper2008,Vos2017,Suarez2022,Suarez2023}). It is also vital for understanding the differences between free-floating planetary mass objects and directly imaged exoplanets. Directly imaged exoplanets occupy a unique color-magnitude sequence (e.g., \citealt{Gratton2024}), and the discovery of extremely red, faint objects such as CWISE J0506+0738 (a strong candidate member of the $\beta$ Pictoris Moving Group; \citealt{Schneider2023}), serves to narrow the color-magnitude gap.

In this work, we present VHS J1831$-$5513, an L/T transition dwarf with anomalously red near-infrared colors. At $J-K_{\rm S}=3.633\pm0.277$ mag, it is potentially the reddest known free-floating substellar object, more so than the current reddest spectroscopically confirmed object in the literature, CWISE J0506+0738, by $\sim$0.66 mag. In Section~\ref{sec:ident} we discuss the discovery of VHS J1831$-$5513. In Section~\ref{sec:Observations}, we present Magellan/FIRE spectroscopic follow-up data and in Section~\ref{sec:Analysis}, we analyze and discuss those data. In Section~\ref{sec:Discussion}, we evaluate young moving group membership, estimate physical properties, and investigate the prospects for variability in VHS J1831$-$5513.

\section{Identification of VHS J1831-5513\label{sec:ident}}
The unique color space occupied by young L dwarfs makes it possible to perform targeted searches using a photometric approach; a particularly valuable property given their typically low tangential velocities \citep{Faherty2009,Faherty12}, which makes their discovery through proper motion-based searches (e.g., Backyard Worlds; \citealt{kuchner2017,Humphreys2020}) challenging. Utilizing this principle, we performed a half-sky search of CatWISE2020 (CatWISE; \citealt{Marocco2021}) and VHS for objects with extremely red ($J-K_{\rm S}>2$ mag, $J-W2>4$ mag) near-infrared colors (Bickle et al., in prep). VHS J1831$-$5513 was discovered during a secondary search, which used the CatWISE reject table instead of the main CatWISE catalog, in order to collect any spuriously rejected candidates. VHS J1831$-$5513 was rejected by CatWISE due to the overlap of a latent (a.k.a. charge persistence; \citealt{Eisenhardt2020}) artifact in WISE. We used WiseView\footnote{\url{http://byw.tools/wiseview}} \citep{CaseldenWiseview}, a tool designed to aid the visualisation of motion in WISE time-resolved coadds, to vet VHS J1831$-$5513, and confirmed the presence of an overlapping latent. We therefore checked the unTimely catalog \citep{Meisner2023,Kiwy2022} to investigate if its CatWISE photometry was adversely affected. This showed that both the W1 and W2 bands were affected by $\sim0.1-0.2$ mag in the scan direction where the latent was present, but both to a similar extent, so while the CatWISE W1-W2 color is not completely unaffected, it is approximately accurate, with an offset from the average unTimely W1-W2 color in the epochs where the latent is absent of only 0.012 mag. We also visually inspected the higher-resolution VHS imagery to confirm there were no obvious contaminants or resolved companions affecting its near-infrared colors. Once these checks were complete, and we were satisfied that VHS J1831$-$5513 was a genuine extremely red candidate, it was added to our list for follow-up spectroscopy.

\section{Observations\label{sec:Observations}}
We used the 6.5m Magellan Baade telescope and the Folded-port InfraRed Echellette (FIRE; \citealt{SimcoeFIRE}) spectrograph to obtain near-infrared spectra for VHS J1831$-$5513.  Observations were made on 2023 June 30 under clear conditions. We operated FIRE in the low resolution prism mode using the 0\farcs8 slit  (resolution $\lambda$/$\Delta \lambda \sim$100) covering the full 0.8$-$2.5 $\micron$ wavelength range. The source was observed in an ABBA pattern with 16 individual exposures of 120 seconds each, using the Sample Up the Ramp (SUTR) read mode.  Immediately after the science image we obtained a Neon, Neon, Argon lamp and the A0 star HD 153842 for telluric, wavelength and flux calibration. 
All data were reduced using a custom version of the \texttt{FIREHOSE} \citep{FIREHOSE} package. The pipeline was modified such that boxcar extractions were performed on A-B images to avoid large variations in the slope of the resulting spectra, and subroutines from the SpeXtool v1 package \citep{spextool,Vacca2003} were used to combine, mask, and inspect the resulting spectra.

\begin{figure*}
    \centering
    \includegraphics[scale=0.55,angle=0]{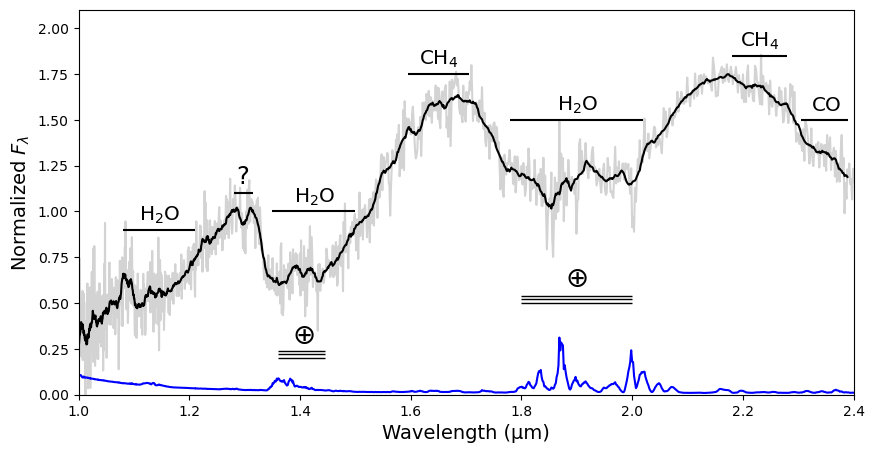}
    \caption{Magellan/FIRE spectrum of VHS J1831$-$5513 (black), Gaussian smoothed using a width of 15 pixels. Key molecular absorption features are labeled. Original resolution data are plotted in grey and flux uncertainty is shown in blue.}
    \label{fig:spectrum}
\end{figure*}

The final reduced spectrum is shown in Figure~\ref{fig:spectrum}, in which prominent absorption features and telluric bands are labeled. We caution that due to the nonlinear response of the detector, the slope after the $\sim$2.3 $\micron$ CO edge may be unreliable as response decreases rapidly towards the edge of the detector.

\section{Analysis\label{sec:Analysis}}

\begin{deluxetable}{lcc}
\label{tab:properties}
\tablecaption{Properties of VHS J1831$-$5513}
\tablehead{
\colhead{Parameter} & \colhead{Value} & \colhead{References}}
\startdata
$\alpha$ [J2000] (\degr) & 277.898262 & 1\\
$\delta$ [J2000] (\degr) & $-$55.232219 & 1\\
$\mu_{\alpha}$\tablenotemark{a} (mas yr$^{-1}$) & 1.43$\pm9.6$ & 2\\
$\mu_{\delta}$\tablenotemark{a} (mas yr$^{-1}$) & $-$84.24$\pm10.4$ & 2\\
\textit{J} (mag) & 20.175$\pm0.229$ & 1\\
\textit{$K_{s}$} (mag) & 16.542$\pm0.048$ & 1\\
W1 (mag) & 14.735$\pm0.018$ & 2\\
W2 (mag) & 13.926$\pm0.016$ & 2\\
\textit{d}$_{\text{phot}}$\tablenotemark{b} (pc) & 55.4$\pm5.5$ & 3\\
Spectral Type [NIR] & L8$\gamma$-T0$\gamma$ & 3
\enddata
\vspace{5pt}
\textbf{Notes}:
\tablenotetext{a}{Proper motion values corrected for CatWISE2020 systematic offsets.}
\tablenotetext{b}{Photometric distance estimated using the VHS \textit{$K_{s}$} magnitude and the absolute magnitude-spectral type relation of \cite{DupuyLiu2012}.}
\tablerefs{(1) VHS DR6 \citep{mcmahon2013} (2) CatWISE2020 Reject Table \citep{Marocco2021} (3) This work}
\end{deluxetable}

\subsection{Spectral Type}
\begin{figure*}
    \centering
    \includegraphics[scale=0.57,angle=0]{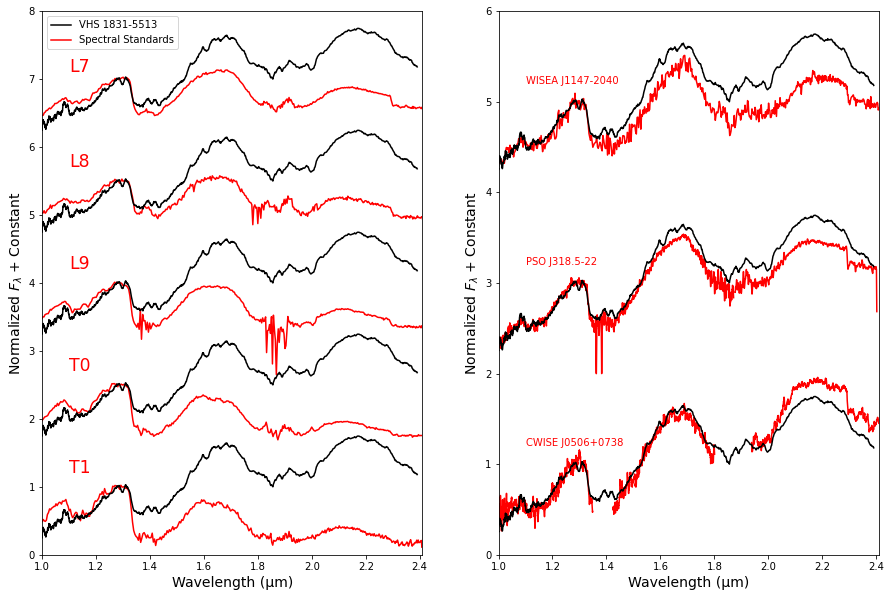}
    \caption{Left: Smoothed Magellan/FIRE spectrum of VHS J1831$-$5513 (black) compared to the spectral standards (red, labeled) 2MASS J08251968+2115521 (L7) from \cite{Cruz2018}, 2MASS J16322911+1904407 (L8) and 2MASS J02550357$-$4700509 (L9) from \cite{Kirkpatrick2010}, and SDSS J120747.17+024424.8 (T0) and SDSS J083717.21$-$000018.0 (T1) from \cite{Burgasser2006}. Right: VHS J1831$-$5513 compared to the young, planetary-mass objects WISEA J1147$-$2040 \citep{Schneider2016}, PSO J318.5$-$22 \citep{Liu2013} and CWISE J0506+0738 \citep{Schneider2023}. All spectra are normalized between 1.27 and 1.29 $\micron$.}
    \label{fig:standardscomp}
\end{figure*}

The left panel of Figure~\ref{fig:standardscomp} compares the near-infrared spectrum of VHS J1831$-$5513 to several field-age L/T transition spectral standards \citep{Burgasser2006,Kirkpatrick2010,Cruz2018}. The extremely red spectral slope and low gravity features (e.g., steep \textit{H}-band slope) in the near-infrared spectrum of VHS J1831$-$5513 leads to poor fits to the spectral standards over the 1--2.4$\micron$ range as a whole, as is the case for many other young, red objects (e.g., \citealt{Faherty2013,Liu2013,Schneider2016,Schneider2023}). The best fit over the $J$-band is the L9 standard 2MASS J02550357$-$4700509 \citep{Kirkpatrick2010}. Although progress is being made in the spectral classification of young L and T dwarfs (e.g., \citealt{Piscarreta2024}), there is not yet a unified approach tailored to young objects, nor a complete set of young L/T transition spectral standards. The diversity seen in the near-infrared slopes and molecular absorption features within the known young populations remains a challenge for the development of a uniform classification technique of young objects. The right panel of Figure~\ref{fig:standardscomp} compares VHS J1831$-$5513 to the young, red objects WISEA J114724.10$-$204021.3 (WISEA J1147$-$2040; L7$\gamma$; \citealt{Schneider2016}), PSO J318.5338$-$22.8603 (PSO J318.5$-$22; L7 VL-G; \citealt{Liu2013}) and CWISE J050626.96+073842.4 (L8-T0$\gamma$; \citealt{Schneider2023}). These comparisons highlight the extremely red near-infrared slope of VHS J1831$-$5513, which is noticeably redder than many members of the young planetary-mass population, and is matched only by CWISE J0506+0738, the reddest free-floating L/T dwarf currently known.

\begin{figure*}
    \centering
    \includegraphics[scale=0.6,angle=0]{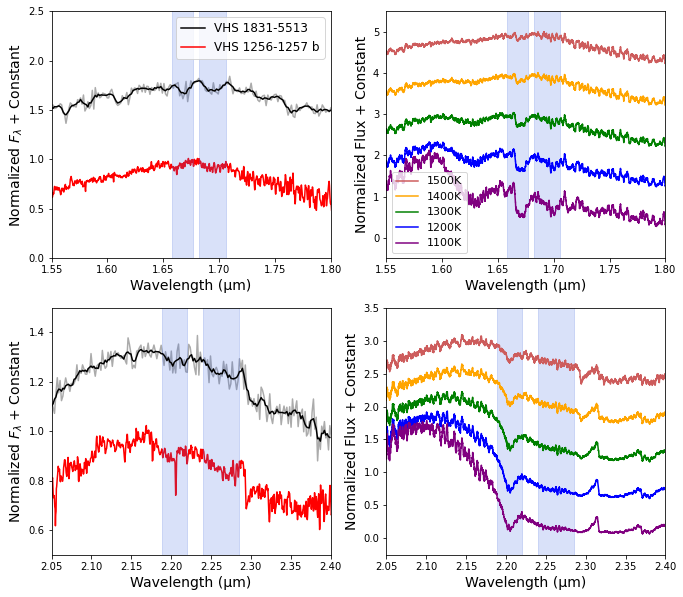}
    \caption{Left column: Comparison between the \textit{H}- (top) and \textit{K}- (bottom) bands of VHS J1831$-$5513 (black - smoothed; grey - original resolution) and the young, planetary-mass companion VHS 1256-1257 b (red; \citealt{Miles2023}). CH$_4$ absorption features that appear in both objects are marked by blue bands. Right column: Sonora Bobcat \citep{Marley2021} models at solar metallicity and log(\textit{g}) = 3.5, with a range of effective temperatures. The blue bands represent the locations of the absorption features in the corresponding band of the left column.}
    \label{fig:vhs1256b}
\end{figure*}

Like CWISE J0506+0738, the spectrum of VHS J1831$-$5513 displays likely CH$_{\rm 4}$ absorption features at the \textit{H}-band peak and on the red side of the \textit{K}-band peak. These features also occur in the spectrum of another young, planetary-mass L dwarf, VHS 1256$-$1257 b. The left column of Figure~\ref{fig:vhs1256b} compares the \textit{H}- and \textit{K}-bands of the Magellan/FIRE spectrum of VHS J1831$-$5513 and the \textit{James Webb Space Telescope} (JWST) spectrum of VHS 1256$-$1257 b \citep{Miles2023}. The two objects have several likely CH$_{\rm 4}$ absorption features in common, as highlighted in the left column of Figure~\ref{fig:vhs1256b}. The wavelengths of these features coincide with CH$_{\rm 4}$ absorption bands that begin to form in model spectra of low gravity objects at effective temperatures $\lesssim1500$~K and become more prominent as the temperature decreases. The right column of Figure~\ref{fig:vhs1256b} shows Sonora Bobcat \citep{Marley2021} models at solar metallicity and a constant surface gravity (log(\textit{g})$=3.5$), with temperatures ranging from 1100K to 1500K. Highlighted are the locations of the absorption features that occur in the spectra of both VHS J1831$-$5513 and VHS 1256$-$1257 b. 

Unlike CWISE J0506+0738 and VHS 1256$-$1257 b, VHS J1831$-$5513 displays a prominent absorption feature at the peak of the \textit{J}-band, centered at $\sim$1.3 $\micron$. This feature is not reproduced by the Sonora models and, to our knowledge, no feature centered at this wavelength has previously been noted in other L dwarfs. In T dwarfs, a 1.3 $\micron$ CH$_{\rm 4}$ band is responsible for the gradually increasing downward slope on the red side of the \textit{J}-band peak, starting in early T dwarfs and progressing through later subtypes. It eventually merges with the 1.4 $\micron$ H$_{\rm 2}$O band to produce the characteristic, thin \textit{J}-band peak of late-T dwarfs \citep{Burgasser2002}. Given the presence of likely CH$_{\rm 4}$ absorption elsewhere in the spectrum of VHS J1831$-$5513, the observed 1.3 $\micron$ feature may also be the result of CH$_{\rm 4}$ absorption. Given that such absorption has only been seen in T dwarfs, this would be suggestive of VHS J1831$-$5513 being narrowly on the T side of the L/T transition. However, we caution that CH$_{\rm 4}$ is not known to exhibit any sharp features in this area, and this would therefore be a very unusual presentation for CH$_{\rm 4}$ absorption. There is a known telluric O$_2$ feature close to the wavelength in question (though not at the exact wavelength) which may cause contamination (\citealt{Kausch2015} and references therein). We therefore inspected the raw science and telluric data to check if the feature was the result of poor telluric correction, but no obvious cause was found. The feature also does not appear in any of the other spectra taken with the instrument on the night. For now, we are uncertain of the feature's cause, and characterize it as an unknown absorber. Future acquisition of higher resolution, higher signal-to-noise (S/N) spectra will allow this feature to be investigated further using the spectral inversion (retrieval) technique, which has been effectively utilized in recent studies to constrain cloud properties and elemental abundances of substellar objects (e.g., \citealt{Burningham2021,Calamari2022,Vos2023,Hood2023,Adams2023,Whiteford2023}).

Considering the L9 standard being the best \textit{J}-band fit, the deep 1.1~$\micron$ H$_{\rm 2}$O band, and the \textit{H}- and \textit{K}-band CH$_{\rm 4}$ absorption features, we assigned a preliminary spectral type of L8-T0 (v. red), prior to gravity assessment.

\subsection{Gravity}
Although the extremely red near-infrared colors of VHS J1831$-$5513 are a potential indicator of youth, a small population of field gravity objects are also known to possess red colors (e.g., \citealt{Marocco2014}), possibly due to an excess of sub-micron sized dust grains in their photospheres \citep{Hiranaka2016}. A near equator-on inclination angle has also been found to be a cause of red colors among field objects \citep{Vos2017,Toro2022,Suarez2023}, indicating that red colors in brown dwarfs are more complex diagnostically than previously thought. Hence, other spectral diagnostics must be performed to confirm youth. 

Youth in brown dwarfs and planetary-mass objects can also be inferred through evidence of low surface gravity, as young objects are still contracting \citep{Burrows1997} and have larger radii than equivalent temperature field objects. Intermediate and low surface gravity objects, suffixed $\beta$ and $\gamma$ respectively (\citealt{Kirkpatrick2005}; or INT-G and VL-G per \citealt{AllersLiu2013} indices), can be discerned from field gravity objects by utilizing morphological differences in their optical and near-infrared spectra. Gravity-sensitive indices are well established for early-to-mid type L dwarfs \citep{Lucas2001,Mcgovern2004,Kirkpatrick2006,Cruz2009,AllersLiu2013,Martin2017,Cruz2018}. Some rely on the weakening of atomic lines in the spectra of lower gravity objects. Others measure the shape of spectral regions, the morphologies of which are particularly affected by the strength of H$_2$ collision-induced absorption (CIA), which is reduced in lower gravity objects. However, most gravity-sensitive indices do not function well for late type L dwarfs ($>$L5).

One index that may function for objects as late in type as VHS J1831$-$5513 is the \textit{H}-slope index \citep{Schneider2023}. Low gravity (young) objects have more peaked \textit{H}-bands caused by reduced H$_2$ CIA in their low-pressure photospheres. \textit{H}-slope aims to quantify this by using a linear least-squares fit between 1.45-1.64 $\micron$ to measure the slope of the blue side of the \textit{H}-band, where the unit is normalized flux/wavelength. \cite{Schneider2023} found that for L7-T0 dwarfs, typical field gravity objects have \textit{H}-slope values ranging from 2-4, and low gravity objects have values between 3-5. VHS J1831$-$5513 has an \textit{H}-slope value of 5.068, considerably higher than those of field objects, and even slightly exceeding the top end of the young population. This result is extremely similar to that of the spectrum of CWISE J0506+0738, which returns an \textit{H}-slope value of 5.043\footnote{\cite{Schneider2023} reported an \textit{H}-slope value of 4.38 for CWISE J0506+0738, calculated before rescaling to correct for Keck/NIRES interband flux calibration issues. Our value for CWISE J0506+0738 is calculated using its spectrum rescaled to its photometric $J-K$ color.}.

The H$_2$(\textit{K}) index \citep{Canty2013} has also been shown to differentiate low gravity and field gravity objects at late-L subtypes \citep{Schneider2014}. Like the \textit{H}-cont and \textit{H}-slope indices, it utilizes the reduced H$_2$ CIA that occurs in lower gravity objects, but in the \textit{K}-band. To do this, it measures the slope between 2.17 $\micron$ and 2.24 $\micron$ by calculating the ratio of the median fluxes in 0.02 $\micron$ ranges centered at those wavelengths in the unsmoothed spectrum of an object. For L7-T0 spectral types, typical index scores for field gravity objects are $1.05\leqslant$H$_2$(\textit{K})$\leqslant1.2$, whereas low gravity objects score much lower than their field gravity counterparts of the same spectral type \citep{Schneider2014}. VHS J1831$-$5513 returns a score of H$_2$(\textit{K})=1.045, over 1$\sigma$ below the field sequence at L7, and significantly further below for later spectral types ($\sim$3$\sigma$ below for L8), where the field sequence sees a rapid increase (see fig. 10 of \citealt{Schneider2014}).

We caution that as most of these indices predate the discovery that inclination angle is correlated with cloud opacity \citep{Vos2017,Toro2022,Suarez2023}, they are not designed to account for this, and as such, it is possible that the viewing geometry of VHS J1831$-$5513 may interfere with the gravity assessment they provide. The existence of such contamination and extent to which this may impact the index scores is yet to be fully tested.

Considering the extremely red near-infrared slope, the \textit{H}-slope and H$_2$(\textit{K}) indicators, and the probable kinematic match to the $\sim$22 Myr $\beta$ Pictoris Moving Group \citep{Zuckerman2001,Shkolnik2017} discussed later, we conclude that VHS J1831$-$5513 has low surface gravity, and tentatively classify it as L8$\gamma$-T0$\gamma$. Both a higher S/N spectrum and the future development of gravity-sensitive indices applicable to L/T transition objects will contribute to a higher confidence gravity assessment for VHS~J1831$-$5513.

\section{Discussion\label{sec:Discussion}}

\subsection{Moving Group Membership\label{sec:YMG}}
Young moving groups typically retain similar 3D velocities for a few hundred Myr, and their population encompasses the whole range of masses from planetary-mass to stars \citep{ZuckSong2004,Torres2008,Malo2013,Gagne2014,Faherty2016}. Young moving groups are coeval, so objects that are found to be a member of a group can be assigned the age of that group and become key benchmarks for its population.

To determine if VHS J1831$-$5513 is a member of a known young moving group, we used the BANYAN $\Sigma$ algorithm \citep{Gagne2018}, which uses Bayes' theorem to select the best-matching XYZUVW model among 27 young associations and the distribution of nearby field stars, while optionally marginalizing over missing radial velocities or distances with an exact, analytical solution to the marginalization integrals. BANYAN $\Sigma$ requires minimum inputs of position and proper motions, and can optionally use distance and heliocentric radial velocity with respective uncertainties. If distance and/or radial velocity are not provided, it calculates and provides optimal values for these, assuming membership in each of the 27 groups. 

The long time baseline of CatWISE (12 epochs over 8 years) means that it can provide proper motion measurements with reasonable confidence ($\pm\sim$10 mas yr$^{-1}$ for a W$1=14$ mag source; \citealt{Marocco2021}). We therefore use the CatWISE proper motion data for the BANYAN~$\Sigma$ input. However, CatWISE has known systematic astrometric offsets\footnote{{\url{https://irsa.ipac.caltech.edu/data/WISE/CatWISE/gator_docs/CatWISE2020_Table1_20201012.tbl}}} compared to Gaia DR2. We therefore applied the offsets for the CatWISE tile in which VHS J1831$-$5513 is located, and used the corrected proper motion values for our BANYAN~$\Sigma$ input. When only provided with the position and proper motions of VHS J1831$-$5513, with respective uncertainties, BANYAN~$\Sigma$ gives a 79.7\% membership probability for the $\sim$22 Myr $\beta$ Pictoris Moving Group (BPMG; \citealt{Zuckerman2001,Shkolnik2017}), an 8.6\% probability for the $\sim$45 Myr Argus Association (ARG; \citealt{Torres2008,Zuckerman2019}), a 2\% probability for the $\sim$130 Myr AB Doradus Moving Group (ABDMG; \citealt{Zuckerman2004,Gagneabdmgage}), and a 9.8\% probability of VHS J1831$-$5513 not being a member of any known group.

VHS J1831$-$5513 is too faint at optical wavelengths to be detected by the \textit{Gaia} mission \citep{Gaiadr3}, and in surveys where it \textit{is} detected, insufficient astrometric data of adequate precision to determine a parallax exist. We instead calculated a photometric distance for VHS J1831$-$5513. As has been noted in previous works, objects with unusual near-infrared colors do not conform well to standard absolute magnitude relations \citep{Faherty2013,Faherty2016,Filippazzo2015,Liu2016}. In investigating this, \cite{Schneider2023} compared the parallactic distances of several very red, free-floating objects to their \textit{J}, \textit{K}, W1, and W2 photometric distances, and showed that for those objects, the \textit{K}-band photometric distance provided the best match on average to the measured parallax. We therefore used the VHS \textit{K$_s$}-band magnitude to calculate a photometric distance for VHS J1831$-$5513 using the absolute magnitude-spectral type relations in \cite{DupuyLiu2012}. Using a spectral type of L9$\pm1$, we determine a \textit{K$_s$}-band photometric distance for VHS J1831$-$5513 of 55.4$\pm5.5$ pc. This distance agrees well with the optimal kinematic distances for BPMG (51.5$\pm5.1$ pc) and ARG (54.8$\pm5.8$ pc) membership provided by BANYAN $\Sigma$. Adding this photometric distance to the existing position and proper motion parameters, BANYAN $\Sigma$ gives an 85.2\% membership probablity for BPMG, 9.7\% for ARG, 0.4\% for ABDMG and 4.6\% for no group. While promising, this is not a conclusive indicator of BPMG membership, as a probability of $>90\%$ is typically used as a cutoff to select high probability group members \citep{Gagne2015}. Regardless, the indications of low surface gravity in its spectrum and the fairly low precision proper motion measurements from CatWISE for VHS J1831$-$5513 merit further investigation of a $\sim85\%$ membership probability. A summary of the BANYAN $\Sigma$ results are provided in Table~\ref{tab:YMG}. 

In light of its positive spectral indicators of youth and its best BANYAN $\Sigma$ kinematic match, we find that VHS J1831$-$5513 is most likely a member of the BPMG. A higher precision proper motion measurement, as well as future radial velocity and trigonometric parallax measurements are required to confirm or refute this membership.

\begin{deluxetable*}{lcc}
\label{tab:YMG}
\tablecaption{BANYAN $\Sigma$ Membership Results}
\tablehead{
\colhead{Input Data} & \colhead{Group Membership Probability (\%)\tablenotemark{a}} & \colhead{d$_{opt}$ (pc)\tablenotemark{b}}}
\startdata
$\mu$ & BPMG (79.7), ARG (8.6), ABDMG (2), Field (9.8) & BPMG (51.5$\pm$5.1), ARG (54.8$\pm$5.8), ABDMG (67.1$\pm$6)\\
$\mu$ + d$_{phot}$ & BPMG (85.2), ARG (9.7), ABDMG (0.4), Field (4.6) & -
\enddata
\vspace{5pt}
\textbf{Notes}:
\tablenotetext{a}{ABDMG = AB Doradus Moving Group \citep{Zuckerman2004}; ARG = Argus Association \citep{Torres2008}; BPMG = $\beta$ Pictoris Moving Group \citep{Zuckerman2001}; Field = Not a member of any known group.}
\tablenotetext{b}{Optimal distance to VHS J1831$-$5513, assuming membership of individual groups.}
\end{deluxetable*}

\subsection{Physical Properties}\label{sec:PhysPro}

Under the assumption that it is a member of the BPMG, certain physical properties of VHS J1831$-$5513 can be inferred. \cite{Viana2009} found that the BPMG has approximately solar metallicity, so we assume solar metallicity for our analysis. Firstly, we estimated the bolometric luminosity of VHS J1831$-$5513. To do this, we used the \cite{Sanghi2023} empirical bolometric correction-spectral type polynomial relations for young objects, in combination with the absolute VHS $K_{\rm S}$-band magnitude based on our previously calculated photometric distance. We estimate a bolometric luminosity for VHS J1831$-$5513 of log$(L_{bol}/L_{\odot})=-4.57\pm0.09$. Note that several assumptions are made when determining this value, such as our photometric distance being accurate and the \cite{Sanghi2023} relations being applicable to such an unusually red object. A more accurate bolometric luminosity can be determined with a measured parallax and broader wavelength coverage.

To test this value, we also derive a bolometric correction for VHS 1256-1257b using the bolometric luminosity from \cite{Miles2023} and photometry from the Ultracoolsheet (\citealt{ultracoolsheet} and references therein), and apply it to VHS J1831$-$5513. VHS 1256-1257b is known to be red and young, and the broad wavelength coverage provided by the JWST observations of this object allows for a precise determination of its L$_{\text{bol}}$ value. This yields a value of log$(L_{bol}/L_{\odot})=-4.51\pm0.09$ for VHS J1831$-$5513.

We used the Sonora Bobcat \citep{Marley2021} and Diamondback \citep{Morley2024} evolutionary models at solar metallicity to estimate the mass and radius of VHS J1831$-$5513 using the age of BPMG ($22\pm6$ Myr; \citealt{Shkolnik2017}) and the bolometric luminosity derived from the \cite{Sanghi2023} relations. Using the Sonora Bobcat models, assuming BPMG membership, we infer a radius of $1.32^{+0.03}_{-0.02}$ $R_{\text{Jup}}$ and a mass of $7\pm1$ $M_{\text{Jup}}$. Using the calculated bolometric luminosity, derived radius, and the Stefan-Boltzmann law, we estimate an effective temperature for VHS J1831$-$5513 of $T_{\text{eff}}=1130\pm60$ K. From the Sonora Diamondback models, taking the same age and bolometric luminosity, we infer a radius of $1.42^{+0.04}_{-0.03}$ $R_{\text{Jup}}$, a mass of $6.5\pm1.5$ $M_{\text{Jup}}$ and an effective temperature of $T_{\text{eff}}=1085\pm60$ K.

While both models suggest a similar mass and effective temperature for VHS J1831$-$5513, the radius is notably different. The Diamondback values are preferred, as the model incorporates silicate clouds, as well as gravity-dependent cloud clearing, which appears to occur at the L/T transition \citep{Marley2010}. In both cases, the mass is well below the deuterium burning limit of $\sim$13 $M_{\text{Jup}}$ (\citealt{Spiegel2011} and references therein) and the effective temperature is $\sim$150-200 K cooler than an L9 at field age \citep{Kirkpatrick2021}, which is consistent with previous findings that young objects are cooler than field age objects of equivalent spectral type \citep{Filippazzo2015,Faherty2016,Liu2016,Suarez2021}. An insufficient number of L/T transition objects are currently known in BPMG to infer the temperature at which the L/T transition occurs in the group. However, the temperature at which the L/T transition occurs decreases with age, and \cite{GagneTransition} found that the L/T transition for ABDMG is $\sim$1150 K, suggesting that if VHS J1831$-$5513 is a true member of BPMG, the L/T transition may occur at a similar temperature in the two groups, despite the $\sim$100 Myr age difference.

VHS J1831$-$5513 is notably $\sim$100 K cooler than PSO J318.5$-$22 \citep{Miles2018} which displays no corresponding near-infrared CH$_{4}$ absorption features. It is however, an extremely similar effective temperature to CWISE J0506+0738 ($T_{\text{eff}}=1140\pm80$ K; \citealt{Schneider2023}), which exhibits likely CH$_{4}$ absorption at both the \textit{H}- and \textit{K}-band peaks. This temperature similarly bolsters the empirical temperature constraint provided by CWISE J0506+0738 for the introduction of CH$_{4}$ absorption into the near infrared spectra of young L/T transition objects.

BANYAN $\Sigma$ returns a non-zero probability of VHS J1831$-$5513 being a member of ABDMG, which has very similar UVW and distance values among its membership to that of BPMG. These similarities are known to confuse membership assignment for some objects between the two groups \citep{Gagne2018}. In case of the unlikely event that VHS J1831$-$5513 is an ABDMG member, we repeated the analysis using the age of the ABDMG ($133^{+15}_{-20}$ Myr; \citealt{Gagneabdmgage}). Assuming ABDMG membership, the \cite{Marley2021} models suggest a radius of $1.23^{+0.13}_{-0.11}$ $R_{\text{Jup}}$, an effective temperature of $1170\pm50$ K and a mass of $12\pm0.2$ $M_{\text{Jup}}$ for VHS J1831$-$5513. This would still place VHS J1831$-$5513 below the deuterium burning limit, solidifying its status as a planetary-mass object. We reiterate that the scenario where VHS J1831$-$5513 is a member of ABDMG is unlikely, and BPMG is the strong favorite for membership.

\subsection{Prospects for Variability}\label{sec:Var}

Young L and T dwarfs are known to exhibit stronger photometric and spectrophotometric variability than field age objects, driven by rotational modulations of condensate clouds in their atmospheres \citep{Biller2015,Biller2018,Metchev2015,Lew2016,Schneider2018,Vos2018,Vos2019,Vos2020,Vos2022,Eriksson2019,MilesPaez2019,Bowler2020,Manjavacas2021}. Furthermore, the L/T transition has been shown to have a higher variability occurrence rate than other spectral types \citep{Radigan2014}. Indeed, \cite{Liu2024} determined that the variability rate of young objects at the L/T transition is $64^{+23}_{-22}\%$, compared to $31^{+12}_{-9}\%$ outside the L/T transition. Variable L dwarfs are more likely to have cloudier atmospheres \citep{Suarez2022} and redder colors \citep{Ashraf2022} than non-variable counterparts. Additionally, both variability amplitude and infrared colors have been found to be correlated with inclination angle, with equator-on objects having higher amplitudes and redder colors than those which are viewed pole-on \citep{Vos2017}. Near-infrared colors also depend on cloud thickness, where redder L dwarfs have higher cloud opacity than bluer objects \citep{Suarez2022}. 

\cite{Suarez2023} further demonstrated a correlation between infrared color, cloud opacity, and viewing inclination for L dwarfs, finding that equator-on objects exhibit more clouds and, therefore, are redder than objects viewed close to pole-on. This indicates that equatorial latitudes are cloudier than polar latitudes and the color diversity of brown dwarfs is strongly influenced by their viewing geometry.

\begin{figure*}
    \centering
    \includegraphics[scale=0.31,angle=0]{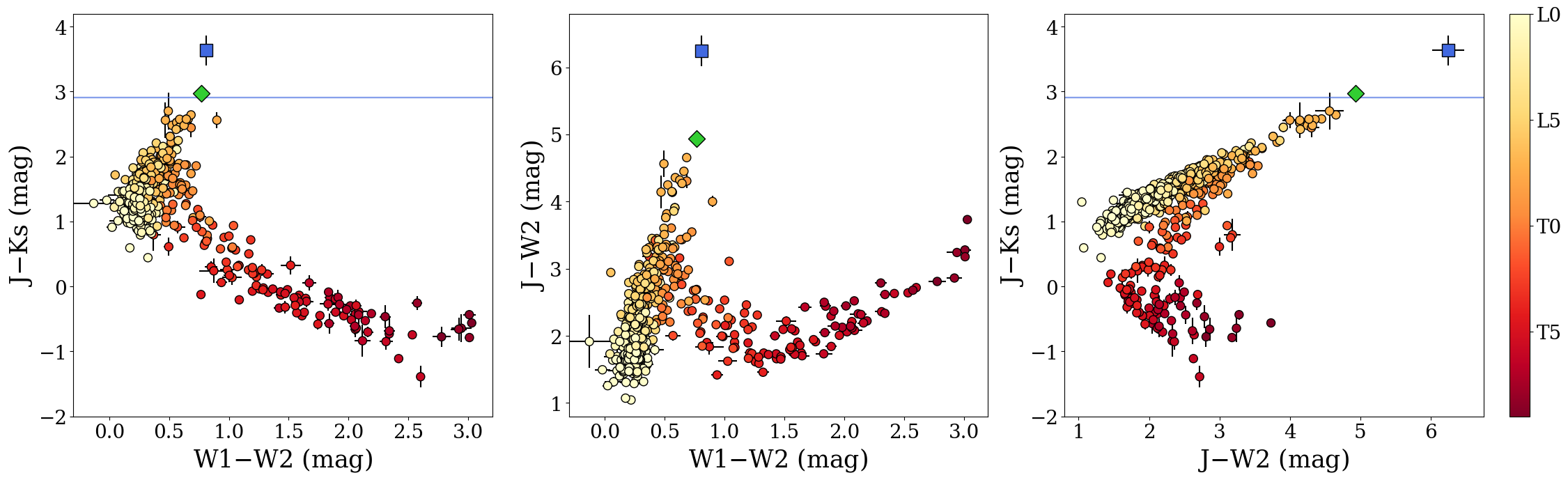}
    \caption{Color-color diagrams comparing VHS J1831$-$5513 (blue square) to L and T dwarfs with photometry from \cite{Schneider_phot} (yellow to red circles) and CWISE J0506+0738 (green diamond; \citealt{Schneider2023}). VHS J1831$-$5513 is plotted using its VHS and CatWISE photometry, and represents a significant outlier from even the reddest known objects. The blue lines in the left and right plots indicate the synthetic $J-K_{\rm S}$ color calculated from the Magellan/FIRE spectrum of VHS J1831$-$5513, which is still an outlier from the bulk of the brown dwarf population, but is similar to that of CWISE J0506+0738, the reddest known L/T transition dwarf.}
    \label{fig:colorcolors}
\end{figure*}

The VHS photometry of VHS J1831$-$5513 indicates a color of $J-K_{\rm S}=3.633\pm0.277$ mag. However, visual inspection and comparison of its Magellan/FIRE spectrum showed that its spectrum is not as red as CWISE J0506+0738 (see Figure~\ref{fig:standardscomp}), which has a color of $(J-K)_{\rm MKO}$=2.974$\pm0.039$ mag \citep{Schneider2023}. We therefore extracted synthetic photometry from the spectrum of VHS J1831$-$5513 using a custom code with filter profiles from SVO \citep{SVO2012,SVO2020}. This gives a color of $(J-K_{\rm S})_{\text{synth}}=2.914\pm0.008$ mag, which is discrepant from the VHS color by 0.719 mag (3.07$\sigma$). The photometric and synthetic colors of VHS J1831$-$5513 are compared to the photometric colors of CWISE J0506+0738 and other L and T dwarfs in Figure ~\ref{fig:colorcolors}.

The VHS \textit{J}- and $K_{\rm S}$-band exposures were taken $\sim$6 minutes apart, so can be treated as near-simultaneous; the shortest known variability period of a brown dwarf is $\sim$1.08 hr \citep{Tannock2021,Vos2022}, and young brown dwarfs are known to be slow rotators \citep{Schneider2018,Vos2022}. Thus, such a short time difference should be insufficient to cause a significant enough photometric change to explain the observed color discrepancy in VHS J1831$-$5513.

While we cannot definitively rule out that the sole cause of the color discrepancy is the low S/N of the VHS \textit{J}-band detection (S/N$\approx$4.6), we note that VHS J1831$-$5513 possesses exceptionally red near-infrared colors, which is suggestive of an extreme atmosphere, potentially viewed near equator-on. Complex clouds, disequilibrium chemistry, and powerful atmospheric dynamics may be expected in such a young, red atmosphere \citep{Showman2013,Suarez2023b,Suarez2023}, and can affect the emergent flux observed from the object. It is therefore possible that cloud-driven variability in VHS J1831$-$5513 may contribute to the observed $(J-K_{\rm S})$ color discrepancy. The $\sim$9 yr difference between the VHS images being taken and the Magellan/FIRE spectrum being collected is easily sufficient for variability to have played a role in causing the discrepancy between the two.

Pressure plays a key role in the behavior of clouds within the atmospheres of substellar objects \citep{Buenzli2012,Yang2016}. With the \textit{J}- and $K_{\rm S}$-bands probing different atmospheric depths, they may exhibit different variability amplitudes, as has been observed in other L/T transition objects, which show higher amplitudes in the \textit{J}-band than the $K_{\rm S}$-band \citep{Artigau2009,Radigan2012,Vos2019}. VHS 1256$-$1257 b, an object similar in nature to VHS J1831$-$5513, is highly variable in the \textit{J}-band, with both \cite{Bowler2020} and \cite{Zhou2020} finding a 1.27 $\micron$ variability amplitude of $\sim$25$\%$, and \cite{Zhou2022} finding a broader \textit{J}-band variation of up to $38\%$ on a $\sim$1 year timescale, the highest ever observed in a substellar object. In the same vein, phase shifts between \textit{J}- and $K_{\rm S}$ light curves can also affect the $J-K_{\rm S}$ color of L/T transition dwarfs (e.g., \citealt{Mccarthy2024,Plummer2024}). Future time-resolved monitoring will provide insight into whether differing variability amplitudes and/or a phase shift between the \textit{J}- and \textit{K$_{\rm S}$}-bands of VHS J1831$-$5513 exist, and if so, to what extent they impact its $J-K_{\rm S}$ color.

The correlation between near infrared color, viewing inclination, and cloud opacity in \cite{Suarez2023} is helpful to understand the appearance of objects like VHS J1831$-$5513. This may indicate that its extremely red near-infrared colors and potentially the CH$_{4}$ features in its Magellan/FIRE spectrum, are explained, at least partially, by unusually thick clouds as a consequence of the object being viewed equator-on. We note, however, that the \cite{Suarez2023} study focused on L3--L7 dwarfs and did not take into account objects as late in type as VHS J1831$-$5513. L7 marks the peak for $J-K$ colors on the field sequence as objects at later types rapidly get bluer as condensates sink below the photosphere. While the trends between near-infrared color and inclination in \cite{Vos2017} and \cite{Vos2020} are observed into the L/T transition, it is an evolutionary stage in which marked changes, both chemical and physical, occur in the atmospheres of brown dwarfs, so further investigation is needed to determine if the trends identified in \cite{Suarez2023} are applicable to the L/T transition.

The only multi-epoch survey to detect VHS J1831$-$5513 is the WISE mission. However, because of the latent artifact contamination described in Section \ref{sec:ident}, we were unable to confidently explore the variability properties of VHS J1831$-$5513 using the WISE data. Nonetheless, VHS~1831$-$5513 is highly likely to exhibit detectable variability given its red infrared colors, young age, possible near equator-on inclination angle and cloudy atmosphere, and its position at the L/T transition \citep{Radigan2014,Vos2017,Vos2019,Suarez2022}, making it a promising target for future variability studies.

\section{Conclusion}
We have presented the discovery of VHS J1831$-$5513, an extremely red, young L/T transition object, identified through a search of the CatWISE reject and VHS catalogs. We obtained Magellan/FIRE near-infrared spectroscopy of VHS J1831$-$5513, which shows that it is an $\sim$L8$\gamma$-T0$\gamma$ dwarf with likely CH$_{4}$ absorption at the \textit{H}- and \textit{K}-band peaks. The presence of CH$_{4}$ in the atmosphere of VHS J1831$-$5513 indicates a $T_{\rm eff}$ at the L/T transition.  We find that its spectral morphology is similar to those of other young, red late-type L dwarfs, including the planetary mass companion VHS 1256$-$1257 b, with which it shares a number of individual CH$_{4}$ absorption features. VHS J1831$-$5513 also exhibits a 1.3 $\micron$ absorption feature, the cause of which is unclear. This feature must be investigated further, potentially at higher resolution, in order to confidently assert its cause.

Using the BANYAN $\Sigma$ Bayesian algorithm, we determine that VHS J1831$-$5513 is a likely member of the $\sim$22 Myr BPMG based on its kinematics. This membership assignment is supported by its low gravity score in pressure-sensitive indices and its red near-infrared slope. Future trigonometric parallax, radial velocity, and higher precision proper motion measurements will help to test this membership. Using the age of BPMG, we inferred physical properties for VHS J1831$-$5513. Its estimated mass of $6.5\pm1.5$ $M_{\text{Jup}}$ places it among the lowest mass free-floating BPMG members known (alongside CWISE J0506+0738 at $7\pm2$ $M_{\text{Jup}}$; \citealt{Schneider2023}), contributing to empirical constraints on the group's initial mass function. Its estimated effective temperature of $T_{\text{eff}}=1085\pm60$ K is well below that of field-age objects of similar spectral types, consistent with previous findings for other young brown dwarfs. Given the spectroscopic evidence of the presence of CH$_{4}$ in its atmosphere, this temperature estimate also places empirical constraints on the temperature at which CH$_{4}$ begins to appear in the near-infrared spectra of young L/T transition dwarfs.

The extremely red near-infrared colors and potentially near equator-on inclination angle of VHS J1831$-$5513 make it highly likely to be extremely cloudy. This, in combination with its young age, and position at the L/T transition, suggests that VHS J1831$-$5513 is highly likely to be variable, and is therefore an excellent prospective target for future photometric and spectrophotometric monitoring. A significant ($\sim$3.1$\sigma$) $J-K_{\rm S}$ color discrepancy noticed between its Magellan/FIRE spectrum and VHS photometry may also indicate that it has a significant difference between its \textit{J}- and $K_{\rm S}$-band variability amplitudes or a potential phase shift between its \textit{J}- and $K_{\rm S}$ light curves, making it a yet more intriguing target for variability monitoring.

The discoveries of VHS J1831$-$5513 in this work and CWISE J0506+0738 in \cite{Schneider2023} hint at the existence of a population of exceptionally red L/T transition objects which, until the introductions of newer near-infrared surveys with deeper detection thresholds, were simply too faint to be detected in the \textit{J}-band by the previous generation of near-infrared survey instruments (e.g., 2MASS). Future targeted searches of these catalogs are warranted to discover more examples of such extraordinarily red objects. This also highlights the importance of the upcoming Euclid mission \citep{EuclidSurvey}, which has an expected 5$\sigma$ point source \textit{J}-band depth of $\sim$24.5 mag \citep{EuclidNISP}, and so will be able to detect not only intrinsically fainter red objects, but red objects at further distances, expanding the census of known red objects and enabling a better understanding of their space density, a key component for exploring the mass functions of nearby associations.

VHS J1831$-$5513 is an excellent prospective JWST target to facilitate atmospheric retrieval studies with high resolution data. This would also represent an excellent opportunity to probe the L/T transition at young ages; an area not yet fully understood, primarily due to the paucity of known young L/T transition objects.

\acknowledgments
\noindent

The Backyard Worlds: Planet 9 team would like to thank the many Zooniverse volunteers who have participated in this project. We would also like to thank the Zooniverse web development team for their work creating and maintaining the Zooniverse platform and the Project Builder tools. This research was supported by NASA grant 2017-ADAP17-0067. This material is supported by the National Science Foundation under grant No. 2007068, 2009136, and 2009177. JF acknowledges the Heising Simons Foundation as well as NSF award \#1909776, and NASA Award \#80NSSC22K0142. J. M. V. acknowledges support from a Royal Society - Science Foundation Ireland University Research Fellowship (URF$\backslash$1$\backslash$221932). This research made use of the Montreal Open Clusters and Associations (MOCA) database, operated at the Montr\'eal Plan\'etarium (J. Gagn\'e et al., in preparation). This paper includes data gathered with the 6.5 meter Magellan Telescopes located at Las Campanas Observatory, Chile. This publication makes use of data products from the {\it Wide-field Infrared Survey Explorer}, which is a joint project of the University of California, Los Angeles, and the Jet Propulsion Laboratory/ California Institute of Technology, funded by the National Aeronautics and Space Administration. CatWISE was funded by NASA under Proposal No. 16-ADAP16-0077 issued through the Astrophysics Data Analysis Program, and uses data from the NASA-funded WISE and NEOWISE projects. This work uses data obtained as part of the VISTA Hemisphere Survey, ESO Progam, 179.A-2010 (PI: McMahon). This research has made use of the Spanish Virtual Observatory (https://svo.cab.inta-csic.es) project funded by MCIN/AEI/10.13039/501100011033/ through grant PID2020-112949GB-I00. This work has benefitted from The UltracoolSheet at http://bit.ly/UltracoolSheet, maintained by Will Best, Trent Dupuy, Michael Liu, Aniket Sanghi, Rob Siverd, and Zhoujian Zhang, and developed from compilations by Dupuy \& Liu (2012, ApJS, 201, 19), Dupuy \& Kraus (2013, Science, 341, 1492), Deacon et al. (2014, ApJ, 792, 119), Liu et al. (2016, ApJ, 833, 96), Best et al. (2018, ApJS, 234, 1), Best et al. (2021, AJ, 161, 42), Sanghi et al. (2023, ApJ, 959, 63), and Schneider et al. (2023, AJ, 166, 103).

\facilities{Magellan/FIRE, Paranal/VISTA, WISE}

\software{\texttt{FIREHOSE} \citep{FIREHOSE}, SpeXtool v1 \citep{spextool}, unTimely Catalog Explorer \citep{Kiwy2022}, WiseView \citep{CaseldenWiseview}}

\bibliography{paper}
\bibliographystyle{aasjournal}

\end{document}